\documentclass[12pt, a4paper]{article}
\usepackage{amsmath,amssymb}
\usepackage{graphicx}
\usepackage{color}
\usepackage{cite,fancyhdr}
\usepackage[affil-it]{authblk}

\begin{document}

\begin{titlepage}

\begin{center}

\hfill UT-17-20, TU-1045,  IPMU17-0077\\

\vskip .75in

{\Large \bf 
Neutrino CP phases from Sneutrino Chaotic Inflation
}

\vskip .75in

{\large Kazunori Nakayama$^{a,b}$, Fuminobu Takahashi$^{c,b}$ and Tsutomu T. Yanagida$^{b,d}$}

\vskip 0.25in

\begin{tabular}{ll}
$^a$ &\!\! {\em Department of Physics, Faculty of Science, }\\
& {\em The University of Tokyo,  Bunkyo-ku, Tokyo 113-0033, Japan}\\[.3em]
$^{b}$ &\!\! {\em Kavli IPMU (WPI), UTIAS,}\\
&{\em The University of Tokyo,  Kashiwa, Chiba 277-8583, Japan}\\[.3em]
$^{c}$ &\!\! {\em Department of Physics, Tohoku University, }\\
& {\em Sendai, Miyagi 980-8578, Japan}\\
$^{d}$ &\!\! {\em Hamamatsu Professor}
\end{tabular}

\end{center}
\vskip .5in

\begin{abstract}
We study if the minimal sneutrino chaotic inflation is consistent with 
a flavor symmetry of the Froggatt-Nielsen type, 
to derive testable predictions on the Dirac and Majorana CP violating phases, $\delta$ and $\alpha$.
For successful inflation, the two right-handed neutrinos,
i.e., the inflaton and stabilizer fields,  must be degenerate in mass.
First we find that the lepton  flavor symmetry structure becomes less manifest 
in the light neutrino masses in the seesaw mechanism, and this tendency becomes
most prominent when right-handed neutrinos are degenerate. 
Secondly, the Dirac CP phase turns out to be sensitive to whether the shift symmetry breaking depends on 
the lepton flavor symmetry. When the flavor symmetry is imposed only on the stabilizer Yukawa couplings,
distributions of the CP phases are peaked at $\delta \simeq \pm \pi/4, \pm 3\pi/4$ and $\alpha = 0$, while
the vanishing and maximal Dirac CP phases are disfavored.
On the other hand, when the flavor symmetry is imposed on both the inflaton and stabilizer
Yukawa couplings, it is rather difficult to explain the observed neutrino data, and those parameters
consistent with the observation prefer the vanishing CP phases $\delta = 0, \pi$ and $\alpha = 0$.
\end{abstract}

\end{titlepage}


\renewcommand{\thepage}{\arabic{page}}
\setcounter{page}{1}
\renewcommand{\thefootnote}{\#\arabic{footnote}}
\setcounter{footnote}{0}

\newpage

\section{Introduction}

The chaotic inflation~\cite{Linde:1983gd} is attractive in that it naturally avoids the initial 
condition problem of the inflation. In addition it generically predicts a large
tensor-to-scalar ratio within the reach of the ongoing and future CMB B-mode search
experiments. A simple prescription for realizing the chaotic inflation in supergravity
was given in Ref.~\cite{Kawasaki:2000yn}, where two gauge singlets are required for
successful inflation;
one is the inflaton and the other is the stabilizer field. See also 
Refs.~\cite{Kallosh:2010ug,Nakayama:2013jka,Nakayama:2013txa,Nakayama:2014wpa} 
for variants of the inflation models along the same line. 

Recently we proposed a minimal sneutrino chaotic inflation model~\cite{Nakayama:2016gvg},  
where the two gauge singlets, inflaton and stabilizer fields, are  identified with the
two right right-handed neutrinos. In particular, the two right-handed neutrinos are  almost
degenerate in mass, and the mass scale is fixed to be about $10^{13}$\,{\rm  GeV} 
by the CMB normalization~\cite{Ade:2015lrj}.
Interestingly, the seesaw mechanism~\cite{Yanagida:1979as,GellMann:1980vs,Minkowski:1977sc}
as well as leptogenesis~\cite{Fukugita:1986hr} are naturally implemented in the minimal chaotic
sneutrino inflation (see Ref.~\cite{Bjorkeroth:2016qsk} for a recent detailed analysis on the leptogenesis
in this setup).
While the quadratic chaotic inflation is in a strong tension with the observation, 
there are various extensions with  a polynomial 
potential~\cite{Nakayama:2013jka,Nakayama:2013txa,Nakayama:2014wpa,Saha:2015afa},
a periodic potential~\cite{Czerny:2014wza,Czerny:2014xja,Czerny:2014qqa,Nakayama:2016gvg},  a modified (running)
kinetic term~\cite{Takahashi:2010ky,Nakayama:2010kt,Nakayama:2010ga,Kallosh:2013yoa,Kallosh:2016sej},  which give
a better fit to the observed data.
See also Refs.~\cite{Murayama:1992ua,Ellis:2003sq, Kadota:2005mt,Antusch:2009ty,Nakayama:2013nya,
Murayama:2014saa,Evans:2015mta,Antusch:2008pn,Antusch:2010va} for  various inflation models with right-handed sneutrinos.

It is known that, in the seesaw mechanism, the number of parameters
in high energy is larger than that in low energy. For instance, in the conventional case with
three right-handed neutrinos, there are $18$ and $9$ parameters in high and low energy,
respectively.  In a case of two right-handed neutrinos, they are reduced to $11$ and $7$, respectively. 
In the minimal sneutrino chaotic inflation, there are only $9$ 
parameters in high energy, because the right-handed neutrino masses are almost degenerate,
and  fixed to be about $10^{13}$\,GeV by inflation.
The remaining two degrees of freedom correspond to a complex variable $z$ which parametrizes
a complex orthogonal matrix in the Casas-Ibarra parametrization of the neutrino Yukawa matrix~\cite{Casas:2001sr,Ibarra:2003up}.  As we shall see, the imaginary component of $z$ is tightly constrained by the requirement of 
the stability of the inflationary path. We also note that, in the limit of degenerate
right-handed neutrino masses, the real component of $z$ becomes unphysical, further reducing the number
of parameters in high energy.

In this Letter we study if the minimal sneutrino chaotic inflation is consistent with 
a flavor symmetry of the Froggatt-Nielsen (FN)~\cite{Froggatt:1978nt} type, 
to derive testable predictions on the CP phases.
 First we note that, in the seesaw mechanism with degenerate 
right-handed neutrinos, the lepton flavor symmetry structure becomes less manifest in the 
light neutrino masses. 
Secondly, we find that the Dirac CP phase is sensitive to whether the shift symmetry breaking depends on 
the lepton flavor symmetry. The inflaton respects a shift symmetry to ensure the flatness of the inflaton potential
beyond the Planck scale, and so, the inflaton Yukawa couplings (as well as the inflaton mass) explicitly
break the shift symmetry. Since we do not know the origin of the shift symmetry breaking terms, they
may depend on the lepton flavor symmetry.  We therefore study two cases in which the flavor structure is imposed 
on only the stabilzer Yukawa couplings and on both the inflaton and stabilizer Yukawa couplings. In either case,
we randomly scan the full theoretically and observationally allowed parameter space 
of the flavor model, and derive probability distributions of
 the neutrino Dirac and Majorana CP phases, $\delta$ and $\alpha$.

The rest of the Letter is organized as follows. In Sec.~\ref{sec:2} we first review the minimal
sneutrino chaotic inflation, derive constraint on the complex parameter $z$ from the stability
condition of the inflationary path, and explain the flavor model adopted for our analysis.
In Sec.~\ref{sec:3} we  show the probability distributions of the CP phases based on the
random scan of the allowed parameter space. The last section is devoted for discussion and 
conclusions.

\section{Models of inflation and flavor symmetry}
\label{sec:2}

\subsection{Minimal sneutrino chaotic inflation}

The successful chaotic inflation in supergravity requires two gauge singlet superfields, 
the inflaton and stabilizer fields. Being gauge singlets, they are naturally coupled to
leptons and Higgs. Thus, the seesaw mechanism for light neutrino masses is a natural
outcome of the chaotic inflation in supergravity~\cite{Nakayama:2016gvg}.  
Then the inflaton and the stabilizer field are identified with the right-handed sneutrinos. 

The relevant K\"ahler and super-potentials for the sneutrino chaotic inflation are given by
\begin{align}
	K  &= \frac{1}{2}(N_1+N_1^\dagger)^2 + |N_2|^2 - k_2\frac{|N_2|^4}{M_P^2}, \label{Kapp}\\
	W &= M N_1 N_2 + y_{i\alpha} N_i L_\alpha H_u,
	\label{Wapp}
\end{align}
where $N_1$ and $N_2$ are the inflaton and stabilizer fields, respectively, 
$M$ the inflaton mass, $y_{i\alpha}$ the neutrino Yukawa couplings, $L_\alpha$ the
lepton doublet, and $H_u$ the up-type Higgs field. Here $i$ runs over $1$ and $2$,
and $\alpha = e, \mu, \tau$ is the lepton flavor index. We assume that the $N_1$ 
respects a shift symmetry along its imaginary component, which is identified with the inflaton, 
$\varphi \equiv \sqrt{2} {\rm Im} N_1$. The shift symmetry is explicitly broken by both
the inflaton mass $M$ and neutrino Yukawa couplings, $y_{1 \alpha}$.

The inflaton mass is fixed to be $M \simeq 2\times 10^{13}$\,GeV by the CMB normalization,
and so, typical values of $y_{i\alpha}$ is of $\mathcal O(0.1)$ for reproducing the observed neutrino 
masses (see Sec.~\ref{sec:nu}). Then, if we simply extrapolate the above interactions to 
$\varphi \sim \mathcal O(10)M_P$,
the masses of $L_\alpha H_u$ would exceed the Planck mass during inflation.
One can avoid this problem by imposing a discrete shift symmetry on $N_1$ as  in Ref.~\cite{Nakayama:2016gvg}.
Thanks to the discrete shift symmetry, the actual superpotential is modified near the Planck scale
and the masses of $L_\alpha H_u$ become periodic with respect to  $\varphi$.
Thus their masses remain smaller than the Planck mass, in which case we can safely discuss the inflaton dynamics
in the effective field theory. 
In addition, due to this modification, the prediction of the scalar spectral index and the tensor-to-scalar ratio 
can give a better fit to the observation.\footnote{
The quadratic chaotic inflation predicts a too large tensor-to-scalar ratio. To suppress the tensor-to-scalar ratio without changing the prediction of the spectral index significantly, the inflaton potential needs to be flatter and tachyonic (convex up) around $\mathcal O(10)M_P$. This is naturally realized if 
the potential has a periodicity of $\mathcal O(10)M_P$~\cite{Nakayama:2016gvg}.
} 
Alternatively, one may modify the inflaton kinetic term~\cite{Kallosh:2016sej}.
In either case, the right-handed neutrino mass scale remains of order $10^{13}$\,GeV, and
the above forms of the K\"ahler and super-potentials are sufficient for our purpose,  since we are
interested in the observed neutrino parameters in the present vacuum.

\subsection{Seesaw mechanism with two right-handed neutrinos}  \label{sec:nu}

Now let us briefly discuss the seesaw mechanism with two right-handed 
neutrinos~\cite{Frampton:2002qc} based on the superpotential (\ref{Wapp}).
Here and in what follows we work in the charged lepton mass basis.
First we move to the mass eigenbasis in which the right-handed neutrino masses are 
diagonalized (the parameters in this basis are shown with tildes):
\begin{align}
	W = \frac{1}{2}M \widetilde N_i \widetilde N_i + \widetilde y_{i\alpha} \widetilde N_i L_\alpha H_u,
\end{align}
where
\begin{align}
	&\widetilde N_1= \frac{1}{\sqrt 2}(N_1+N_2),~~~~~~\widetilde N_2= \frac{i}{\sqrt 2}(N_1-N_2),\\
	&\widetilde y_{1\alpha}= \frac{1}{\sqrt 2}(y_{1\alpha}+y_{2\alpha}),~~~~~~\widetilde y_{2\alpha}= \frac{i}{\sqrt 2}(-y_{1\alpha}+y_{2\alpha}).
	\label{masseigen}
\end{align}
After integrating out the right-handed neutrinos, we obtain the light neutrino mass matrix as
\begin{align}
	m^\nu_{\alpha\beta} = \frac{v_{\rm EW}^2\sin^2\beta}{M} \widetilde y_{i\alpha}\widetilde y_{i\beta},
	\label{mnu}
\end{align}
where $v_{\rm EW}=174$\,GeV and $\tan\beta \equiv \left<H_u^0\right>/\left<H_d^0\right>$.
It is diagonalized by the Maki-Nakagawa-Sakata (MNS) matrix as
\begin{align}
	m_{\bar \gamma}^\nu \delta_{\bar \gamma \bar \delta} = U_{\bar \gamma\alpha}^{{\rm (MNS)}T} m^\nu_{\alpha\beta} U_{\beta\bar\delta}^{{\rm (MNS)}},
\end{align}
where the subindices with a bar (e.g. $\bar \gamma = 1,2,3$) label the mass eigenstates. 
Note that the lightest neutrino is massless in the two right-handed neutrino scenario.
We take the standard parametrization of the MNS matrix:
\begin{align}
	U_{\beta\bar\gamma}^{{\rm (MNS)}} =\begin{pmatrix}
		c_{12}c_{13} & s_{12}c_{13} & s_{13}e^{-i\delta} \\
		-s_{12}c_{23}-c_{12}s_{23}s_{13}e^{i\delta} & c_{12}c_{23}-s_{12}s_{23}s_{13}e^{i\delta} & s_{23}c_{13} \\
		s_{12}s_{23}-c_{12}c_{23}s_{13}e^{i\delta} & -c_{12}s_{23}-s_{12}c_{23}s_{13}e^{i\delta} & c_{23}c_{13}
	\end{pmatrix}
	\times{\rm diag}
	\begin{pmatrix}
		1 & e^{i\alpha/2} & 1
	\end{pmatrix},
\end{align}
where $s_{ij} = \sin \theta_{ij}$, $c_{ij} = \cos \theta_{ij}$, and $\delta$ and $\alpha$ 
are the Dirac and Majorana phases, respectively.
Note that there are 7 observables in low energy: the two mass eigenvalues, three mixing angles and two CP phases.
On the other hand, the Yukawa matrix $\widetilde y_{i\alpha}$ contains 9 parameters after rotating away the three 
phases of $L_\alpha$ (and the right-handed leptons simultaneously to keep the charged lepton masses unchanged).
In fact, the Yukawa couplings can be expressed in terms of the 7 observables plus one complex parameter. 
One can explicitly solve for the Yukawa matrix as~\cite{Casas:2001sr,Ibarra:2003up}
\begin{align}
	\widetilde y_{i\alpha} = \frac{M^{1/2}}{v_{\rm EW}\sin\beta}R_{i \bar\gamma} \sqrt{m_{\bar \gamma}^\nu}\delta_{\bar\gamma\bar\beta} U_{\bar\beta\alpha}^{\rm (MNS)\dagger}.
	\label{ytilde}
\end{align}
Here $R_{i \bar\gamma}$ is given by
\begin{align}
	R_{i\bar\gamma} = \begin{pmatrix}
		0 & \cos z & -\zeta\sin z \\
		0 & \sin z & \zeta \cos z
	\end{pmatrix}.
\end{align}
for the normal hierarchy and
\begin{align}
	R_{i\bar \gamma} = \begin{pmatrix}
		-\zeta\sin z & \cos z & 0\\
		\zeta \cos z & \sin z & 0
	\end{pmatrix}.
\end{align}
for the inverted hierarchy, where $z$ is an arbitrary complex parameter and $\zeta=+1$ or $-1$.
In the rest of the paper, we  consider only the normal hierarchy because the inverted hierarchy is 
rarely realized in the flavor model described later.

If there are additional information or constraints on the Yukawa matrix, 
one may not be allowed to freely choose $z$ and the CP phases.
Below we consider two such constraints.
One comes from the condition for successful inflation and the other  from flavor symmetry.
Under these conditions, the parameter ranges of $z$ and CP phases are constrained, which enables us to make 
testable predictions on these parameters.

Lastly let us comment on the degenerate limit of the right-handed neutrinos.
Even though the degeneracy is lifted by a small amount in the actual inflation model and 
a small breaking of the degeneracy is required for successful leptogenesis, 
we could study the above mentioned constraints in the limit of degenerate right-handed neutrinos.
In this case, one can further rotate $(\tilde{N}_1,\tilde{N}_2)$ by an arbitrary real orthogonal matrix
to get rid of the real component of $z$, $z_R\equiv {\rm Re}(z)$. In other words,  $z_R$ becomes
unphysical in the degenerate limit. Indeed, we have confirmed that our results do not depend on 
values of $z_R$, and remain valid even if the degeneracy is lifted by a small amount.

\subsection{Stability condition of the inflationary path}

To derive predictions on the CP phases,  we need to somehow constrain the neutrino Yukawa couplings in (\ref{Wapp}).
One important constraint comes from the inflaton dynamics:
for large inflaton field values, there could appear a tachyonic direction 
in the field space of the slepton $L_\alpha$ and Higgs $H_u$~\cite{Nakayama:2016gvg}.  

To see this, let us write down the scalar potential of $L_\alpha$ and $H_u$ during inflation
when $\varphi = \sqrt{2} {\rm Im} N_1$ takes a large field value,
\begin{align}
	V =\left( M\overline{y}_2N_1^* L_2' H_u + {\rm h.c.}\right)
	 +\overline{y}_1^2 |N_1|^2 \left( \left| L_1' \right|^2 + |H_u|^2\right),
\end{align}
where we have set $N_2=0$, which is ensured by the non-minimal K\"ahler potential in (\ref{Kapp}).
 We have also defined
\begin{align}
	&\overline{y}_1 \equiv \sqrt{ \sum_{\alpha} |y_{1\alpha}|^2 },~~~\overline{y}_2 \equiv \sqrt{ \sum_{\alpha} |y_{2\alpha}|^2 }, \label{ybar}\\
	&L_1' \equiv \frac{1}{\overline{y}_1 } \left( \sum_\alpha y_{1\alpha} L_\alpha  \right),
	~~~L_2' \equiv \frac{1}{\overline{y}_2 } \left( \sum_\alpha y_{2\alpha} L_\alpha  \right).
\end{align}
It is clear that $L_1'$ is stabilized at $L_1'=0$ by the huge mass of $\overline{y}_1|N_1|$.
On the other hand, the mass matrix of $(H_u, L_2'^*)$ is given by
\begin{align}
	m^2_{H L_2'} = \begin{pmatrix}
			\overline{y}_1^2 |N_1|^2  & M \overline{y}_2 N_1  \\
			 M \overline{y}_2 N_1^*     &  k H^2
	\end{pmatrix},
\end{align}
where the Hubble mass correction is included for the mass of $L_2'$ with $\mathcal O(1)$ numerical coefficient $k$.
 Note that the Hubble mass correction comes from the quartic coupling in the K\"ahler potential, $\delta K \sim |N_2|^2 |L_2'|^2$.
The coefficient of this coupling is considered to be of order unity but remains undetermined in our setup.
Since $|N_1|$ takes a large value during inflation and $\overline{y}_1$ is 
 sizable for successful seesaw mechanism, the mass eigenvalues during inflation are well approximated by
\begin{align}
	m_{\rm Heavy}^2 \simeq \overline{y}_1^2 |N_1|^2,~~~~~m_{\rm Light}^2\simeq kH^2-\frac{\overline{y}_2^2}{\overline{y}_1^2} M^2.
\end{align}
In order for the inflaton path to be stable,  $m_{\rm Light}^2 > 0$ should hold until the end of inflation.\footnote{
	After inflation ends, $H$ becomes smaller and eventually there appears a tachyonic direction in the zero-temperature potential.
	However, as shown in Ref.~\cite{Nakayama:2016gvg}, the preheating just after inflation is very efficient due to large Yukawa couplings
	and it leads to effective potential that tends to stabilize the slepton and Higgs fields, although a precise calculation is difficult to perform
	due to non-perturbative nature of the particle production.
	We emphasize that, even if a tachyonic direction appears, it is not a serious problem at all: the whole reheating process simply
	becomes a bit more complicated.
}
Noting $H \sim M$ at the end of inflation, we have a constraint from the stability of the inflaton path as
\begin{align}
	k \gtrsim \frac{\overline{y}_2^2}{\overline{y}_1^2}.   \label{k_y1y2}
\end{align}
Since $k$ is considered to be of $\mathcal O(1)$, 
 $\overline{y}_1$ should 
be comparable to $\overline{y}_2$, which is of ${\cal O}(0.1)$. This provides a non-trivial constraint on the Yukawa couplings in the sneutrino 
chaotic inflation model. In particular, it implies that the masses of  $L_\alpha H_u$ would indeed
exceed the Planck mass if one simply extrapolated the inflation model to $\varphi \gtrsim {\cal O}(10)M_P$.
As we pointed out in Ref.~\cite{Nakayama:2016gvg}, 
one solution is to assume that the discrete shift symmetry remains  unbroken, 
which can give a better fit to the observation, while avoiding  the super-Planckian masses
of $L_\alpha H_u$ during inflation.

One can rewrite the stability condition (\ref{k_y1y2}) in terms of the Casas-Ibarra parameter.
By an explicit calculation, we find
\begin{align}
	\overline{y}_1^2 = \frac{M(m^\nu_2+m^\nu_3)}{2v_{\rm EW}^2 \sin^2\beta}e^{-2z_I},~~~~~~
	\overline{y}_2^2 = \frac{M(m^\nu_2+m^\nu_3)}{2v_{\rm EW}^2 \sin^2\beta}e^{2z_I},
\end{align}
where $z_I\equiv{\rm Im}(z)$.
Note here that $m_1^\nu = 0$ in our scenario with two right-handed neutrinos. 
Therefore, $z_I$ can be considered as the order parameter for the shift symmetry breaking 
in the Yukawa sector, because the larger $z_I$ implies
the smaller inflaton Yukawa couplings $|y_{1\alpha}|$, and vice versa. 
Thus the condition (\ref{k_y1y2}) reads
\begin{align}
	k \gtrsim e^{4z_I}.   \label{zI}
\end{align}
Specifically, if we impose $k\lesssim 10$, we must have $z_I \lesssim 0.6$,
while $z_R$ is not restricted by this condition at all. This is consistent with the 
fact that $z_R$ becomes unphysical in the degenerate limit.

\subsection{Flavor symmetry}

Let us now impose a global U(1)$_{\rm FN}$ flavor symmetry of the FN type~\cite{Froggatt:1978nt} 
on the neutrino Yukawa couplings.
We introduce a chiral superfield $\Phi_{\rm FN}$ whose lowest component  develops a non-zero  
vacuum expectation value, $\langle \Phi_{\rm FN}  \rangle= v_{\rm FN}$, leading to
spontaneous break down of the U(1)$_{\rm FN}$. 
The observed mass hierarchy is controlled by the FN suppression factor, 
$\epsilon \equiv v_{\rm FN}/M_P$, where the cut-off scale is set to be the Planck scale.\footnote{
We assume that the U(1)$_{\rm FN}$ is already broken during inflation. To this end one may introduce
another FN field $\bar \Phi_{\rm FN}$ with $Q_{\rm FN} = 1$ to stabilize the FN fields at non-zero values.
Even if the U(1)$_{\rm FN}$ is restored during inflation, our argument is not significantly modified
if $q = 0$. If $q \geq 1$, the neutrino Yukawa couplings could vanish during inflation, and in this case,
there will be no constraint on $z$ from the stability of the inflationary path.
}

The charge assignments are summarized in Table~\ref{FN-charge}, where
we assume that both the inflaton and the stabilzer field are singlet 
under the flavor symmetry, for simplicity. 
First, we choose the FN charges of lepton doublets as $Q_{\rm FN} (L_1) = q+1$, $Q_{\rm FN} (L_2) = q$ and $Q_{\rm FN} (L_3) = q$~\cite{Buchmuller:1998zf,Sato:2000kj} in order to reproduce the observed features of the MNS matrix elements, 
especially the large $\nu_\mu$-$\nu_\tau$ mixing.
 Here $q = 0, 1, 2$ is an integer.
Next we adopt $\epsilon \simeq 0.2$, for which one can approximately explain the observed charged lepton mass hierarchy 
by imposing the FN charge on the right-handed lepton superfields as
$Q_{\rm FN} ({\overline E}_1) = 4$, $Q_{\rm FN} ({\overline E}_2) = 2$, $Q_{\rm FN} ({\overline E}_3) = 0$.
 The charge assignment depends on the precise value of $\epsilon$ and $\tan \beta$.
In the sneutrino chaotic inflation, the gravitino mass needs to be as heavy as $100$\,TeV or heavier to avoid the BBN constraint 
on the gravitino decay~\cite{Murayama:2014saa,Nakayama:2016gvg}, and so, if the sfermion masses are of the same order of the magnitude as the gravitino mass,
a relatively small value of $\tan \beta$ is favored by the observed Higgs boson mass. In this case, one may choose $q \geq  1$.
One can similarly extend the flavor symmetry to 
the quark sector in a way consistent with SU(5) GUT, and then, the FN parameter is directly 
related to the Cabbibo angle. In the following argument, only the FN charges of the lepton doublets
and the right-handed neutrinos are relevant.

\begin{table}[t!]
\begin{center}
\begin{tabular}[t]{|c||c|c|c|c|c|c|}
                    & $L_1$   & $L_2$ & $L_3$  & $N_1$ & $N_2$ &$\Phi_{\rm FN}$  \\
\hline
$Q_{\rm FN}$  &   $q+1$     &   $q$     & $q$   & $0$   & $0$   &$-1$
\end{tabular}
\end{center}
\caption{The FN charge assignment of the lepton doublets and right-handed neutrinos}
\label{FN-charge}
\end{table}%

With the above flavor symmetry, we expect that there are scaling relations among the 
Yukawa couplings of the stabilizer/inflaton field.
Below we consider two cases separately. 
The first case (case 1) is that the only Yukawa couplings of the stabilizer field $N_2$ satisfy the following scaling, 
\begin{align}
|y_{2 e}|:|y_{2\mu}|:|y_{2 \tau}| \simeq \epsilon : 1 : 1.
\label{FNstructure1}
\end{align}
The second case (case 2) is that both the inflaton ($N_1$) and stabilizer ($N_2$) Yukawa couplings satisfy the scaling relation,
\begin{align}
|y_{i e}|:|y_{i\mu}|:|y_{i \tau}| \simeq \epsilon : 1 : 1~~~~~~{\rm for}~~~~ i=1,2.
\label{FNstructure2}
\end{align}
Note that the inflaton Yukawa couplings break the shift symmetry explicitly
and it is not known whether the shift symmetry breaking terms are blind to the lepton flavors or not.
To be general, we study both cases in the next section.

Lastly let us comment on the flavor symmetry in the seesaw mechanism with degenerate right-handed 
neutrino masses. In fact, even if one imposes a certain flavor structure on the neutrino Yukawa couplings,
it becomes less manifest in the low-energy neutrino mass matrix, 
especially when two right-handed neutrino masses are degenerate.
This can be seen by rewriting Eq.~(\ref{mnu}) in terms of the original Yukawa as
\begin{align}
	m^\nu_{\alpha\beta} = \frac{v_{\rm EW}^2\sin^2\beta}{M}\left( y_{1\alpha}y_{1\beta} + y_{2\alpha}y_{2\beta}\right).
\end{align}
In the case 2, we impose  flavor constraints on the absolute magnitudes of the Yukawa couplings, but
their phases are not constrained.
Since the two terms in the parenthesis are of the same order in magnitude but with uncorrelated phases, 
they often add up in a destructive  way.  Specifically, 
diagonal elements (i.e. $m^\nu_{\alpha \alpha}$) tend to be less suppressed compared to off-diagonal ones, because 
the distribution of $|y_{1\alpha}|^2$ is spread more broadly than $|y_{1 \alpha} y_{1 \beta}|$ with $\alpha \ne \beta$
and the cancellation between the two terms occurs less frequently. As a result, 
the off-diagonal elements tend to be slightly suppressed compared to the diagonal ones, 
and the distributions of the mixing angles shift to smaller values.
Thus the flavor structure of $m^\nu_{\alpha\beta}$ is less manifest with respect to that of the Yukawa sector.
In other words, it is difficult for both Yukawa couplings to satisfy the same scaling relation, while keeping the naively expected 
flavor structure of the light neutrino masses, unless the phases of the Yukawa couplings are aligned. 
This feature is most prominent in our present model in which the two right-handed neutrinos are degenerate in mass.
For a general case where the right-handed neutrinos masses are not degenerate, 
there is a similar tendency, but the flavor structure of $m^\nu_{\alpha\beta}$ is more retained, since one of the
Yukawa couplings  tends to give a dominant contribution to each element of the light neutrino mass matrix.

\section{Numerical analysis}
\label{sec:3}

Now let us numerically analyze distributions of the Dirac and Majorana CP phases in the present setup.
We adopt the following values of the mixing angles and mass squared differences
obtained by a global fit to the experimental data~\cite{Esteban:2016qun}:
\begin{align}
\label{nu-fit}
\sin^2 \theta_{12} &= 0.28-0.33,~~~\sin^2 \theta_{23} = 0.39-0.63,
~~~\sin^2 \theta_{13} = 0.020-0.023,\\
\Delta m^2_{21} &= (7.16-7.88) \times 10^{-5}{\rm\, eV}^2,~~~~\Delta m_{31}^2 = (2.44-2.56)
 \times 10^{-3} {\rm\,eV}^2,
\end{align}
where the quoted ranges corresponds to the $2\sigma$ bounds. We generate random numbers of the
mixing angles and mass squared differences within this range. 
The other parameters $z_R$, $z_I$, $\alpha$, and $\delta$ are randomly varied with a
flat prior in the following ranges: 
\begin{align}
&-\pi < z_R < \pi,~~~0 < z_I < 0.6,~~~-\pi \leq \alpha < \pi,~~~-\pi \leq \delta < \pi.
\end{align}
The range of $z_I$ is restricted because of the inflationary stability condition (\ref{zI}).
To take account of both positive and negative branches of the Casas-Ibarra parametrization,
we also choose the sign of $\zeta (= \pm 1)$ randomly. 
Then we can calculate Yukawa matrix through Eq.~(\ref{ytilde}) and compare it with the 
predictions from the flavor structure of the FN model.

\subsection{Case 1: flavor structure only on the stabilizer Yukawas}

First let us study the case 1 in which only the stabilizer Yukawa couplings are required to satisfy the relation (\ref{FNstructure1}).
To ensure the flavor structure of the Yukawa couplings, we define the normalized Yukawa
couplings as
\begin{align}
N \left(\frac{|y_{2e}|}{\epsilon}, ~|y_{2\mu}|, ~|y_{2\tau}| \right),
\end{align}
where $N$ is the normalization factor to set the average of the components equal to unity:
\begin{align}
N \equiv \frac{3}{|y_{2e}|/\epsilon+|y_{2\mu}| +|y_{2\tau}|}.
\end{align}
Then we require that each normalized Yukawa coupling should be within the range of $1\pm \sigma_y$.

\begin{figure}[!t]
\begin{center}
\includegraphics[scale=1.2]{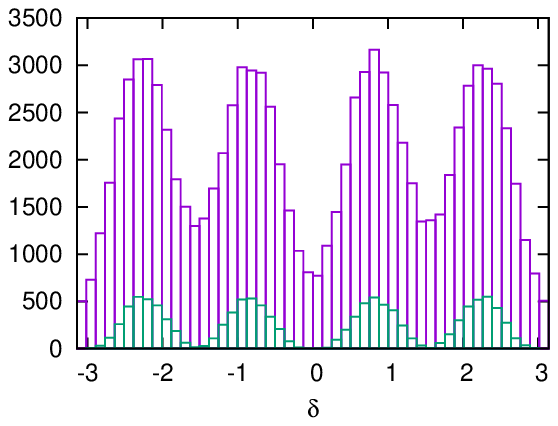}
\includegraphics[scale=1.2]{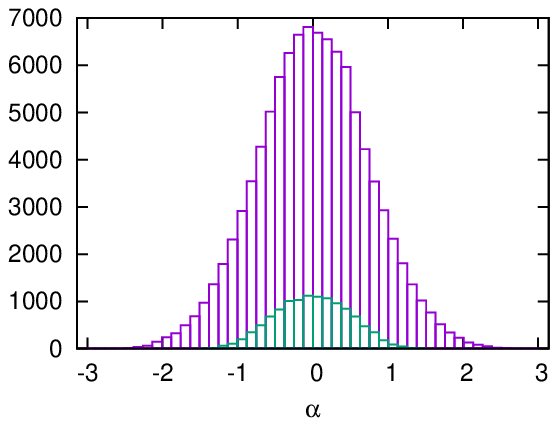}
\caption{
Histogram of the Dirac and Majorana CP phases $\delta$ and $\alpha$ for case 1,
where the flavor structure is imposed only on the stabilizer Yukawa couplings.
We have taken the FN parameter $\epsilon = 0.2$ and set $\sigma_y = 0.2$ (purple)
and $\sigma_y=0.1$ (green).
}
\label{fig:CPphase1}
\end{center}
\end{figure}

In Fig.~\ref{fig:CPphase1} we show the histograms of the Dirac and Majorana CP phases $\delta$ and $\alpha$,
for the FN parameter $\epsilon = 0.2$ with $\sigma_y = 0.2$ (purple) and $\sigma_y=0.1$ (green). 
One can see that the vanishing Dirac CP violation, $\delta = 0$ and  $\pi$,
is disfavored, and that the maximal CP violation, $\delta = \pm \pi/2$, is also (mildly) disfavored. 
The distribution of $\delta$ has peaks located at
$\delta \approx \pm \pi/4$ and $\delta \approx \pm 3\pi/4$. We note that those peaks at $\delta < 0$
correspond to the positive branch $\zeta = 1$ (and vice versa). On the other hand,  
the distribution of the Majorana phase has a broad peak about $\alpha = 0$. 

\begin{figure}[!t]
\begin{center}
\includegraphics[scale=1.2]{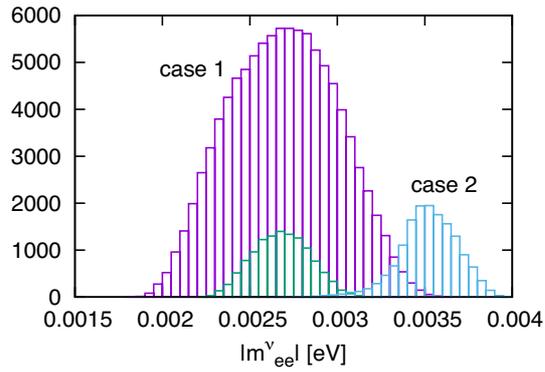}
\caption{
Histogram of $|m_{ee}^\nu|$ in the unit of eV for the case 1 (left: purple and green) and case 2 (right: light blue).
}
\label{fig:mee}
\end{center}
\end{figure}

In Fig.~\ref{fig:mee} we show the histogram of $m_{ee}$ which is relevant for the neutrinoless
double beta decay experiments. 

\subsection{Case 2: flavor structure  on both the inflaton and stabilizer Yukawas}  \label{sec:case2}

\begin{figure}[!t]
\begin{center}
\includegraphics[scale=1.2]{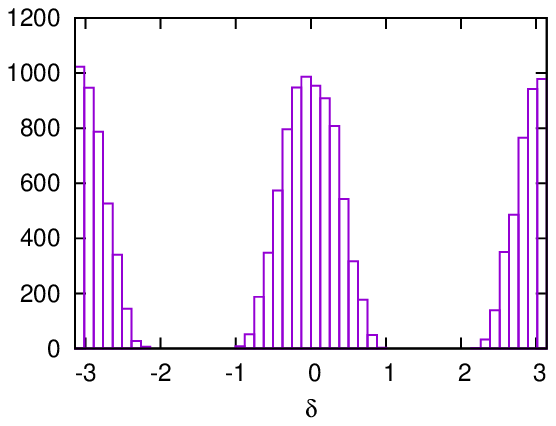}
\includegraphics[scale=1.2]{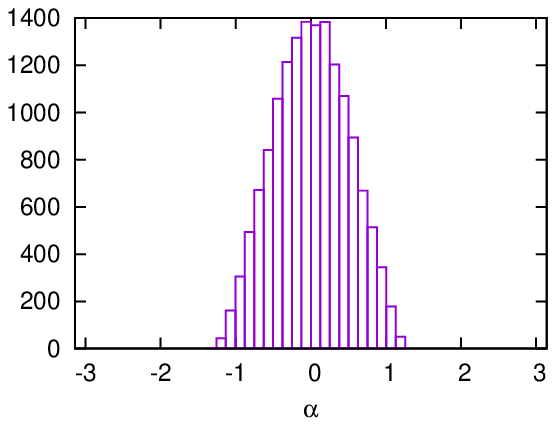}
\caption{
Histogram of the Dirac and Majorana CP phases $\delta$ and $\alpha$ for case 2,
where the flavor structure is imposed on both the stabilizer and inflaton Yukawa couplings.
We have taken the FN parameter $\epsilon = 0.25$ and set $\sigma_y = 0.3$.
}
\label{fig:CPphase2}
\end{center}
\end{figure}

Next let us move to the case 2 in which both the inflaton and stabilizer Yukawa couplings are required 
to satisfy the relation (\ref{FNstructure2}).
Similarly to the previous case, we define the normalized Yukawa couplings as
\begin{align}
N_i \left(\frac{|y_{ie}|}{\epsilon}, ~|y_{i\mu}|, ~|y_{i\tau}| \right),
\end{align}
where 
\begin{align}
N_i \equiv \frac{3}{|y_{ie}|/\epsilon+|y_{i\mu}| +|y_{i\tau}|}
\end{align}
for $i=1,2$.
Then we require that each normalized Yukawa coupling should be within the range of $1\pm \sigma_y$.
We note however that the number of data points which satisfy the scaling relations are much smaller than the previous case.
In fact, with the same FN parameter $\epsilon=0.2$ and the deviation parameter $\sigma_y=0.2$,
we could not find any data point satisfying all the constraints in one million generated data points.
The main reason for the absence of the allowed parameters is that the observed solar mixing angle, $\theta_{12}$, 
is somewhat too large with respect to the expected value in the above flavor model. More precisely, the ratio of the
observed mass squared differences would prefer a much smaller solar mixing angle in the flavor model.
Thus we have relaxed the constraints as $\sigma_y=0.3$ and set $\epsilon=0.25$.
In Fig.~\ref{fig:CPphase2} we show the histograms of the Dirac and Majorana CP phases $\delta$ and $\alpha$ in this case.
One can see that, in contrast to the case 1, the vanishing Dirac CP violation, $\delta = 0$ and $\pi$,
is favored, and that the maximal CP violation, $\delta = \pm \pi/2$, is strongly disfavored. 
On the other hand, the distribution of the Majorana phase has a broad peak about $\alpha = 0$.

As we have seen above,  it is rather difficult to find data points that satisfy all the flavor conditions,
which might imply that it is unlikely that the inflaton Yukawa couplings respect flavor symmetry.
In other words, the shift symmetry breaking terms may not be flavor-blind. In fact, the allowed parameter
region would increase if the inflaton Yukawa couplings had a milder flavor hierarchy with an effective
FN paraemter, $\epsilon_{\rm inf} \sim 0.4 - 0.5$.
This result reflects the tendency that the original flavor structures of Yukawa couplings becomes less manifest
in the low-energy neutrino mass matrix when two right-handed neutrino masses are degenerate.\footnote{
For instance, if one imposes the flavor structure on the light neutrino mass matrix as in Ref.~\cite{Kaneta:2017byo},
the observed solar mixing angle is on the upper end of the distribution. In our case, 
the distribution of the solar mixing angles shifts to smaller values because of the cancellation in the seesaw
formula,  which results in a strong tension between the flavor model and the observation.
}

\subsection{Comparison with previous works}

Before closing, we compare our results with some recent works to clarify the difference.
In Ref.~\cite{Rink:2016knw}, Rink, Schmitz, and one of the present authors (Yanagida) 
studied the two right-handed neutrino scenario with a FN symmetry. They introduced
 the exchange symmetries $\widetilde N_1\leftrightarrow \widetilde N_2$ 
in the right-handed neutrino mass matrix and $\widetilde N_1\leftrightarrow i\widetilde N_2$ in the neutrino 
Yukawa couplings to ensure a certain form of the interactions they considered. 
 They also focused on the case of $|z_I| \gtrsim 2$, which corresponds to 
the so called flavor alignment~\cite{Rink:2016vvl}.
We emphasize here that those symmetries are imposed in the mass eigenstate basis of the right-handed neutrinos.
Under these conditions, they found that the Dirac phase around $\delta \sim -\pi/2$ is favored.
In fact, there are two preferred points around $\delta \sim -\pi/2$, like our result of the case 1 (see Fig.~\ref{fig:CPphase1}).
In our case, the former exchange symmetry is automatically satisfied in the inflation model (\ref{Wapp}).
The latter symmetry is approximately satisfied in our case 1. The reason is as follows. 
Although $z_I$ is bounded above by the inflaton stability condition, 
a mild hierarchy  $|y_{2\alpha}| > |y_{1\alpha}|$ is still possible if $k$ takes a value greater than unity. 
When $|y_{2\alpha}| > |y_{1\alpha}|$, the Yukawa couplings in the mass eigenbasis are
dominated by $y_{2\alpha}$, i.e., $\tilde{y}_{1\alpha} \sim y_{2\alpha}/\sqrt{2}$ and
$\tilde{y}_{2\alpha} \sim i y_{2\alpha}/\sqrt{2}$ (see Eq.~(\ref{masseigen})). Such a structure is also
obtained by imposing the exchange symmetry, $\widetilde N_1\leftrightarrow i\widetilde N_2$.
Therefore, in the limit of flavor alignment, imposing both the flavor structure on $\tilde{y}_{1\alpha}$ and $\tilde{y}_{2\alpha}$ and the exchange 
symmetry $\widetilde N_1\leftrightarrow i\widetilde N_2$ is essentially equivalent to imposing the flavor structure only
on the stabilizer Yukawa coupling, $y_{2\alpha}$. In this sense, the result of  Ref.~\cite{Rink:2016knw}
is roughly consistent with our case 1, although the imposed symmetries are different.

In Ref.~\cite{Kaneta:2017byo}, Kaneta, Tanimoto and one of the present authors (Yanagida)
 studied the distribution of the CP phases in a model with $\det(m^\nu)=0$ motivated by 
 the two right-handed neutrinos, and found a preference for $\delta = \pm\pi/2$.
They randomly scanned parameters in the light neutrino mass matrix, anticipating that
the light neutrino mass matrix satisfies the flavor structure after integrating out the heavy right-handed
neutrinos. We note however that, as we have explained in Sec.~\ref{sec:case2}, 
the flavor structure in the light neutrino mass matrix tends to be less manifest compared to that in
the neutrino Yukawa couplings, because some cancellation among the elements could happen 
in the seesaw formula. In particular, this feature becomes prominent if the right-handed neutrinos are 
degenerate in mass. Therefore, their results can not be directly compared to our case.

\section{Discussion and Conclusions}
\label{discuss}

In the minimal sneutrino chaotic inflation model, the two gauge singlets required 
for successful inflation (i.e. the inflaton and stabilizer fields) are identified with
the right-handed neutrinos, because they are generically coupled to $L_\alpha H_u$~\cite{Nakayama:2016gvg}.
Interestingly, the seesaw mechanism as well as leptogenesis are natural outcomes
of this setup. 

 In this Letter we have investigated this model further, focusing on the structure
of the Yukawa couplings. First, we have derived a constraint on the Yukawa couplings from the stability
condition of the inflationary path, which basically states that the inflaton Yukawa cannot be arbitrarily small
and they are bounded below. Secondly we have imposed a flavor symmetry of the FN type on only the stabilizer Yukawas (case 1) and on both the inflaton
and stabilizer Yukawas (case 2). Under these conditions, we have scanned the parameter space to derive the probability distribution 
of the CP phases, and found that the Dirac CP phase is sensitive to whether the inflaton Yukawa couplings
depend on the lepton flavor symmetry. 
In the case 1, while the vanishing and maximal Dirac CP phases, $\delta = 0,\pi, \pm\pi/2$, are disfavored,
and $\delta \approx \pm \pi/4$ and $\pm 3\pi/4$ are favored.
The distribution of the Majorana CP phase has a broad peak at $\alpha = 0$.
In the case 2, on the other hand, it is hard to explain the observed data based on the flavor model. This is because 
of  the cancellation in the seesaw formula with degenerate right-handed neutrino masses, which results in smaller
mixing angles than naively expected by the flavor symmetry.  If we relax the flavor constraint, it is possible to 
find some parameters consistent with the observation. For those parameters, 
the vanishing Dirac CP phases $\delta = 0,\pi$ are favored and maximal phase $\delta=\pi/2$ is disfavored. The Majorana CP phase has a broad peak at $\alpha = 0$.
We have also clarified differences of our results from those in the literature.

For successful leptogenesis, one needs to lift the degeneracy of the right-handed
neutrino masses. In fact, there is a small contribution to the Majorana mass of $N_1$
from the holomorphic terms $N_1^2 + N_1^{\dag 2}$ in the K\"ahler potential,
which can be absorbed into the superpotential by a K\"ahler transformation~\cite{Bjorkeroth:2016qsk}.
The phase of the Majorana mass comes from the constant term in the superpotential, $W \supset m_{3/2} M_P^2$,
which is needed to realize the vanishingly small cosmological constant. 
Note that the lepton asymmetry is independent of the MNS matrix except for the small flavor effects,
and in particular, those CP phases relevant for leptogenesis are not observed by low-energy experiments.

Let us here point out somewhat peculiar structure of our sneutrino chaotic inflation. First, 
the size of the shift symmetry breaking is quite different in the inflaton mass term
and the neutrino Yukawa couplings. Namely,
while the inflaton mass $M$ is of order $10^{-5}$ in Planck units, the inflaton Yukawa couplings
need to be of order ${\cal O}(0.1)$ due to the stability condition (\ref{k_y1y2}). 
Secondly, we have found that the inflaton Yukawa couplings do not seem to
respect the lepton flavor symmetry, and the observed data implies that the hierarchy in the inflaton Yukawa is milder
than the flavor model predicts. 
Such a peculiar pattern of the symmetry breaking is hard to understand from the low energy point of view. 
However, one may be able to ensure the above features in a set-up with an extra dimension. For instance, let us consider
 a 5D theory compactified on $S_1/Z_2$. We assume that 
 $N_1$, $L_\alpha$, and $H_u$ reside in the bulk and the other fields (including the FN field $\Phi$ as well as
 the stabilizer field $N_2$) are in a brane on one of the boundaries where both the flavor and shift symmetries
 are preserved to a high degree. This ensures the success of the FN model for explaining the charged lepton 
 mass hierarchy as well as the inflaton mass much smaller than the Planck mass. Also the stabilizer Yukawa couplings
 respect the flavor symmetry.
 On the other hand, we assume that 
 the inflaton Yukawa couplings are mainly generated on the other brane where both symmetries are largely broken. This
 explains why the inflaton  has large Yukawa couplings which do not faithfully follow the flavor symmetry.
We note however that one cannot explain why the Majorana mass term of $N_2$, 
 $W = \frac{1}{2}M_2 N_2^2$, is absent or suppressed in the above set-up. This may be explained by the anthropic reasoning: the inflation would 
 not have occurred unless $M_2$ is smaller than the pseudo Dirac mass $M$.
In any case, there may be
 a deep reason for the suggested structure of the inflaton mass and Yukawa couplings, and it is worth studying 
 the model from this perspective.

Finally we point out that the present sneutrino chaotic inflation model with FN symmetry can solve the
cosmological problem of the original model~\cite{Nakayama:2016gvg}.
The problem was that the reheating temperature is as high as $T_{\rm R} \sim 10^{14}$\,GeV
because of the sizable Yukawa couplings of the inflaton/stabilizer and it leads to 
the gravitino overproduction.
In the present model, there is a flavon field whose decay can provide additional entropy production 
to dilute the gravitino abundance.
First note that in the exact global U(1)$_{\rm FN}$ limit, there appears a massless Goldstone boson and 
its scalar partner, sflavon, is also massless up to the soft SUSY breaking~\cite{Ema:2016ops}.
However, we can introduce explicit U(1)$_{\rm FN}$ breaking terms to give these light directions sizable masses $m_\Phi$ $(\ll v_{\rm FN})$.
Therefore we can consider a situation that the sflavon is initially deviated from the minimum in the low energy
by $O(0.1) M_P$ during/after inflation and begins a coherent oscillation when the Hubble parameter 
$H$  becomes equal to $m_\Phi$.
In such a case the (s)flavon dominates the Universe soon after the oscillation and its decay produces a huge amount of entropy.
The (s)flavon may dominantly decay into two leptons plus Higgs, and 
the decay rate is given by $\Gamma_\Phi \sim C\epsilon^\ell m_\Phi^3/M_P^2$ with $C\sim 10^{-4}$ and an integer $\ell$ depending on the detailed FN charge assignments.
The decay temperature is thus becomes $T_\Phi \sim 10^8$\,GeV for $m_\Phi \sim 10^{13}$\,GeV.
(Note that the flavon mass cannot exceed the Hubble parameter during inflation.)
The dilution factor of the preexisting gravitino is about $10^{-2}\,T_\Phi/T_{\rm R}$, which
is enough to dilute the gravitino to a harmless level. The right amount of baryon asymmetry can be generated by
resonant leptogenesis even in the presence of the entropy dilution~\cite{Bjorkeroth:2016qsk}.

\section*{Acknowledgments}

K.N. thanks to So Chigusa for useful discussion.
This work is supported by JSPS KAKENHI Grant Numbers JP15H05889 (F.T.), JP15K21733 (F.T.), 
JP26247042 (F.T),  JP17H02875 (F.T.), JP17H02878(F.T. and T.T.Y),
JP15H05888 (K.N.), JP26800121 (K.N) and JP26104009 (K.N. and T.T.Y), JP16H02176 (T.T.Y),
and by World Premier International Research Center Initiative (WPI Initiative), MEXT, Japan.

\bibliography{reference}

\end{document}